\documentclass[aps,prl,twocolumn,superscriptaddress]{revtex4}
\usepackage{graphicx}

\begin{document}
\preprint{}

\title{Microscopic Evidence for Evolution of Superconductivity by Effective Carrier Doping in Boron-doped Diamond:$^{11}$B-NMR study}

\author{H. Mukuda}
\email[]{e-mail  address: mukuda@mp.es.osaka-u.ac.jp}
\author{T. Tsuchida}
\author{A. Harada}
\author{Y. Kitaoka }
\affiliation{Department of Materials Engineering Science, Osaka University, Toyonaka, Osaka 560-8531, Japan }
\author{\\T. Takenouchi}
\affiliation{School of Science and Engineering, Waseda University, Shinjyuku-ku, Tokyo 169-8555, Japan }
\author{Y. Takano}
\author{M. Nagao}
\author{I. Sakaguchi}
\affiliation{National Institute for Materials Science (NIMS), Tsukuba, Ibaraki 305-0047, Japan }
\author{T. Oguchi}
\affiliation{Department of Quantum Matter, ADSM, Hiroshima University, Higashihiroshima 739-8530, Japan }
\author{H. Kawarada}
\affiliation{School of Science and Engineering, Waseda University, Shinjyuku-ku, Tokyo 169-8555, Japan }

\date{\today}

\begin{abstract}
We have investigated the superconductivity discovered in boron (B)-doped diamonds by means of $^{11}$B-NMR on heteroepitaxially grown (111) and (100) films. $^{11}$B-NMR spectra for all of the films are identified to arise from the substitutional B(1) site as single occupation and lower symmetric B(2) site substituted as boron+hydrogen(B+H) complex, respectively. A clear evidence is presented that the effective carriers introduced by B(1) substitution are responsible for the superconductivity, whereas the charge neutral B(2) sites does not offer the carriers effectively. The result is also corroborated by the density of states deduced by $1/T_1T$ measurement, indicating that the evolution of superconductivity is driven by the effective carrier introduced by substitution at B(1) site. 

%We propose that a possible route to make $T_{\rm c}$  higher in the boron doped diamond is to dissociate the hydrogen at B(2) site, and to make the B(1) concentration increase, keeping the diamond structure. 
\end{abstract}

\pacs{74.78.Db;76.60.Cq;76.60.-k}
%\keywords{superconductivity, diamond, boron, NMR}

\maketitle

%%%%%%%%%%%%%%%%%%%%%%%%%%     Introduction     %%%%%%%%%%%%%%%%%%%%%%%%%%%%

\begin{figure}[htbp]
\begin{center}
\includegraphics[width=0.7\linewidth]{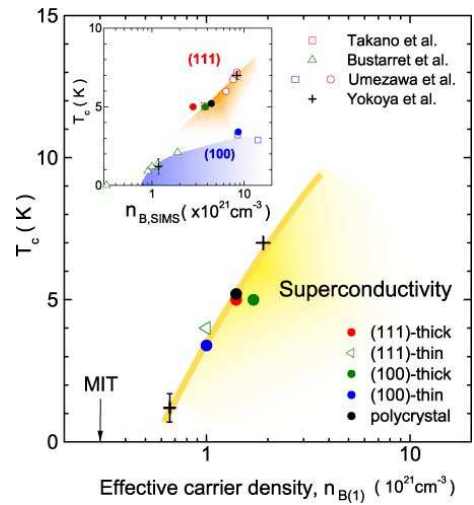}
\end{center}
\caption{(color online)$T_{\rm c}$ is plotted as a function of $n_{\rm B(1)}$, instead of a nominal concentration $n_{\rm B,SIMS}$. A scaling behavior is found between $T_{\rm c}$  and $n_{\rm B(1)}$ corresponding to the effective carrier density. The crosses denote the carrier density estimated from ARPES\cite{Yokoya,Yokoya1}. }
\label{fig:Phase}
\end{figure}

In 2004, Ekimov and coworkers discovered the superconductivity(SC) in heavily boron (B)-doped diamond synthesized by high-pressure and high-temperature technique in very low carrier concentration ($4$-$5\times10^{21}$cm$^{-3}$) \cite{Ekimov}. 
Many theoretical studies have stressed the similarity to MgB$_2$ with a high-$T_{\rm c}$  value of 40 K  where strong coupling of the holes at the top of the valence band with optical phonons plays an important role \cite{Boeri,Lee,Blase}. 
Recently angle-resolved-photoemission spectroscopy (ARPES) revealed that the holes on the diamond bands play an essential role in determining the metallic nature of heavily B-doped diamond superconductors\cite{Yokoya}.
Soon after the first discovery, it has been also reported on the B-doped diamond film synthesized by microwave plasma-assisted chemical vapor deposition (MPCVD) method \cite{Takano}. 
One of the advantages of this method is that B concentration can artificially tune over a wide range, which enables us to elucidate how a superconducting transition temperature ($T_{\rm c}$) depends on a doping level of boron ($n_{\rm B}$) in the homogeneously B-doped diamonds\cite{Bustarret,Umezawa}.
Particularly, Umezawa {\it et al.} have performed the systematic study on homoepitaxially grown (111) and (100) films, and revealed that $T_{\rm c}$ for the (111) films is by more than two times higher than for the (100) films despite of an equivalent B concentration for both films (see inset of Fig.\ref{fig:Phase})\cite{Umezawa}. 
This remarkable difference might contain some hints to understand the mechanism of superconductivity and gives a further insight for searching higher-$T_{\rm c}$  superconductivity in the B-doped diamond and the other related materials. 

In this letter, we report on a microscopic evidence for evolution of superconductivity in B-doped (111) and (100) diamond films, based on the analysis of $^{11}$B-NMR spectra and $^{11}(1/T_1)$ for two predominant boron sites. 

%%%%%%%%%%%%%%%%%%%%%%%%%%     Experimental     %%%%%%%%%%%%%%%%%%%%%%%%%%%%

NMR measurements were performed in five different diamond films; thick and thin (111) heteroepitaxial films, and thick and thin (100) heteroepitaxial films, and thick polycrystalline film, deposited on appropriate substrates for each deposition condition by MPCVD method\cite{Takano,Umezawa}. 
Here the thickness of "thick" and "thin" films in this work are 50$\sim$100$\mu$m and $\sim$3$\mu$m, respectively. 
The nominal B concentration $n_{\rm B,SIMS}$ is derived from secondary ion-mass spectroscopy (SIMS) measurement. 
$T_{\rm c}$  of each sample is determined by an onset of diamagnetism under zero-field cooling, which coincides with an offset of zero resistance. 
NMR spectra were obtained by fast Fourier transform (FFT) technique at a constant magnetic field using a conventional phase-coherent-type NMR spectrometer.  
The Knight shift of $^{11}$B-NMR was determined by the relative shift from the resonance frequency of B(OH)$_3$ (K[B(OH)$_3$]$\approx$ 20 ppm). 
The nuclear spin-lattice relaxation rate $1/T_1$ were measured at a constant magnetic field of 2.5 and 10 T.

%%%%%%%%%%%%%%%%%%%%%%%%%%     Results and Discussions   %%%%%%%%%%%%%%%%%%%

%  1. Spectrum analysis
%**************  Fig.1  *****************************************************
\begin{figure}[htbp]
\begin{center}
\includegraphics[width=0.8\linewidth]{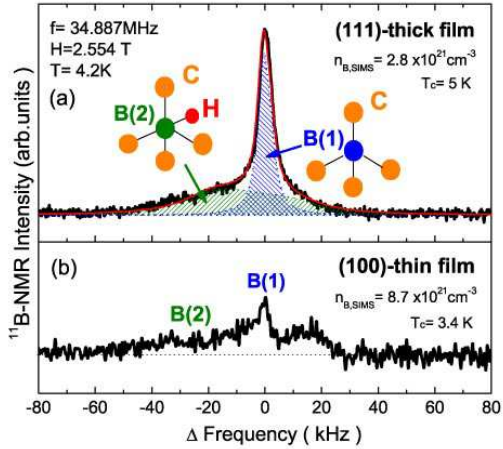}
\end{center}
\caption{(color online)$^{11}$B-NMR spectra for (a) (111)-thick and (b) (100)-thin films, which are typical that $T_c$ for the former is higher than that for the latter, although $n_{\rm B,SIMS}$ for the former is about three times smaller than that for the latter. The calculated spectrum (red curve) for the (111)-thick film is well reproduced by superposition of two components originated from B(1) and B(2) sites.}
\label{fig:spectra}
\end{figure}
%****************************************************************************
%**************  Fig.2  *****************************************************
\begin{figure}[htbp]
\begin{center}
\includegraphics[width=1\linewidth]{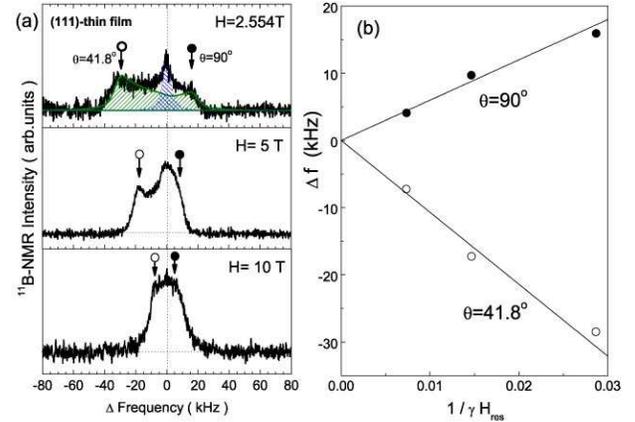}
\end{center}
\caption{(color online) (a)The broad peak becomes narrower in the higher field in all of the films.  (b)The field dependence of the spectral edges is well reproduced by the case of $\nu_Q\sim$1.8MHz as indicated by solid lines.  It provides a clear evidence for presence of EFG only in B(2) site. }
\label{fig:fielddep}
\end{figure}
%****************************************************************************

Figures \ref{fig:spectra}(a) and \ref{fig:spectra}(b) show $^{11}$B-NMR spectra at 4.2 K for the (111)-thick and (100)-thin films, which are noteworthy that $T_c=5$ K for the (111)-thick film is higher than $T_c=3.4$ K for the (100)-thin film, although $n_{\rm B,SIMS}=2.8\times10^{21}$(cm$^{-3}$) for the former is smaller than $n_{\rm B,SIMS}=8.7\times10^{21}$(cm$^{-3}$) for the latter.  
In both films, two spectra overlap with a narrow peak around $\Delta f=0$ with a linewidth of 5 kHz and broad ones in the range $\Delta f=-40$ and $+$20 kHz, respectively. 
For the narrow peak, the Knight shift is approximately -10 ppm, and for the broad one their shifts are distributed over a wider range, both of which do not depend on temperature in the range of 1.4 - 200 K. 
The narrow spectrum originates from the most symmetric B site, i.e., the most likely substitutional position for the carbon site, the linewidth of which should be narrow due to the uniaxial symmetry along each [111] direction at the carbon site in the diamond structure. 
We denote it as B(1) site hereafter. 
On the other hand, the broader spectrum ought to arise from the borons in lower local symmetry, such as boron+hydrogen (B+H) complex site, the interstitial B sites, B+B paired occupation sites, and so on. 
We identify this line-broadening as arising from additional nuclear quadrupole interaction derived from an electric field gradient(EFG) at $^{11}$B nuclear site. 
It is discriminated as B(2) site. 
Generally, in case of $^{11}$B nucleus (I=3/2) where the nuclear quadrupole interaction is sufficiently smaller than Zeeman interaction, the spectrum for resonance peak(I=+1/2$\Leftrightarrow$-1/2) are broaden by the second order perturbation of EFG, which is given by $\Delta f(\theta)=3\nu_Q^2\sin^2\theta(1-9\cos^2\theta)/(16\gamma_N H)$, where $\nu_Q$ is nuclear quadrupole frequency that is proportional to EFG, $\gamma_N$ is a nuclear gyromagnetic ratio and $\theta$ is an angle between the principal axis of EFG and external field that should be taken as a random distribution in this material. 
The right and left edges given when $\theta=90^\circ$ and $\theta=41.8^\circ$ follow the relation of $+3\nu_Q^2/(16\gamma_N H)$ and $-\nu_Q^2/(3\gamma_N H)$ against $H$, respectively. 
Actually, the spectral edges of the broad peak indicated by arrows in Fig. \ref{fig:fielddep}(a) become narrower in the higher field, the field dependence of which is well reproduced by the case of $\nu_Q\approx $1.8MHz for  (111)-thin film, as indicated by solid lines in Fig. \ref{fig:fielddep}(b). 
This behavior has been seen in all of the films in our study, evidencing the presence of EFG only for B(2) site.
Among those possible lower symmetric B sites, we consider that the B substitution for carbon site as B+H complex is most likely than others taking account of the MP-CVD process using the mixed gas of CH$_4$, (CH$_3$)$_3$B and H$_2$\cite{broadpeak}, as discussed also by {\it ab initio} calculation\cite{Yoshida,Oguchi}. 
In this case, the presence of positively charged hydrogen makes the spectrum broader in association with EFG as discussed above, and does not offer the carriers effectively since the positively charged hydrogen compensates the carrier doped by B substitution. 
In fact, the recent SIMS measurement in these films has revealed the presence of higher hydrogen density for lower $T_{\rm c}$ films\cite{Takenouchi}.
In this context, the narrow and broad NMR spectra are identified as arising from the substitutional B site as single occupation (B(1) site) and as B+H complex (B(2) site), respectively, and are totally reproduced by the composition of two different spectra from B(1) site and B(2) site with reasonable parameters\cite{parameters}, as shown in Fig. \ref{fig:spectra}(a). 

\begin{figure}[htbp]
\begin{center}
\includegraphics[width=0.7\linewidth]{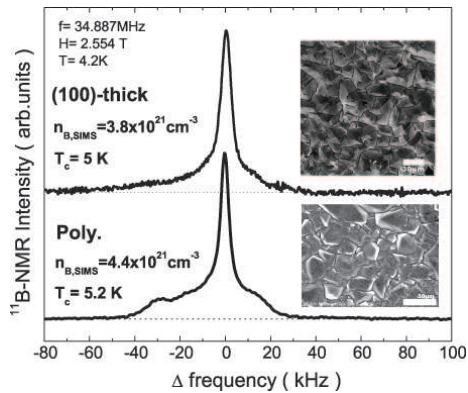}
\end{center}
\caption{(color online)NMR spectra and the SEM images of surface for (100)-thick and polycrystalline films. These spectral shapes seem similar to that of (111)-thick film, because both surfaces are dominated by the \{111\} growth facet.}
\label{fig:SEM}
\end{figure}

It is remarkable that the intensity of the B(1) site is predominantly observed for the (111)-thick film, whereas not for the (100)-thin film. 
Since the NMR intensity is proportional to the number of B nuclei, we can estimate a fraction of B(1) site ($f_{\rm B(1)}$) against nominal B density, by means of evaluating a fraction of the narrow spectrum against an intensity integrated over a whole spectrum. 
The estimated $f_{\rm B(1)}$'s are about 49\% for (111)-thick film and 12\% for (100)-thin film. 
With respect to the (100)-thick film ($n_{\rm B,SIMS}=3.8\times10^{21}$cm$^{-3}$, $T_c\sim$5 K) and the polycrystalline films ($n_{\rm B,SIMS}=4.4\times10^{21}$cm$^{-3}$, $T_c\sim$5.2 K), the NMR spectral shapes are somewhat similar to that of (111)-thick film, as shown in Fig. \ref{fig:SEM}.  
The surface observed by the scanning electron microscopy (SEM) are found to be not flat any longer due to the presence of the preferential \{111\}-growth facets. 
The $f_{\rm B(1)}$'s are estimated to be $\sim$33\% and $\sim$44\% for the polycrystalline and the (100)-thick films, respectively. 
%**************  Fig.3  *****************************************************
%****************************************************************************

The B-concentration dependence of $T_{\rm c}$  are quite important to unravel an origin for the onset of superconductivity in the B-doped diamond. 
We propose that $T_{\rm c}$  should be plotted as a function of the boron concentration at B(1) site ($n_{\rm B(1)}$) that is determined by $n_{\rm B,SIMS}\times f_{\rm B(1)}$, instead of a nominal concentration $n_{\rm B,SIMS}$. 
As shown in Fig. \ref{fig:Phase}, it has been found that $T_{\rm c}$ 's in several different grown films are plotted on a rather simple curve as a function of $n_{\rm B(1)}$. 
This scaling behavior presents clear evidence that effective carriers introduced by the B substitution as B(1) site are responsible for the onset of superconductivity. 
On the other hand, the result also demonstrates that the B(2) site with B+H complex does not contribute to superconductivity because of the substitution as a neutral charged state. 
It is also corroborated that the effective carrier density estimated by ARPES\cite{Yokoya,Yokoya1} is also plotted on the extrapolated curve in Fig. \ref{fig:Phase}. 
%**************  Fig.4 *****************************************************
%****************************************************************************

%**************  Fig.5 *****************************************************
\begin{figure}[htbp]
\begin{center}
\includegraphics[width=0.7\linewidth]{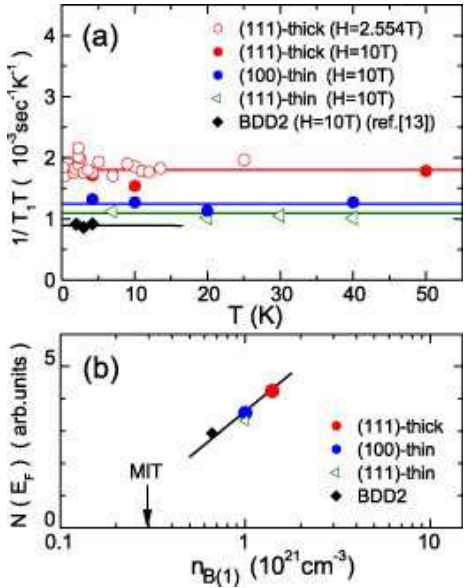}
\end{center}
\caption{(color online) (a)$T$-dependence of $1/T_1T$  for (111) and (100) films. (b)$N(E_{\rm F})$ deduced by $\sqrt{(1/T_1T)}$ are plotted against effective carrier density $n_{\rm B(1)}$.  (111)-thick film with higher $n_{\rm B(1)}$ has larger $N(E_{\rm F})$ than (100)-thin film with lower $n_{\rm B(1)}$ does, suggesting that the enhancement of $T_{\rm c}$ in (111)-thick film relates to the increase of $N(E_{\rm F})$. }
\label{fig:T1}
\end{figure}
%****************************************************************************

This result is also corroborated by  measurements of the nuclear spin-lattice relaxation rate $1/T_1$ for (111)-thick and (100)-thin films. 
As shown in Fig. \ref{fig:T1}(a), $1/T_1T$ stays constant against $T$, obeying so-called "Korringa's relation" over a whole $T$-range, demonstrating that SC is realized under the Fermi liquid state for both films despite the vicinity of metal-insulator-transition(MIT) $n_c\approx 3\times10^{20}$cm$^{-3}$\cite{MIT}. 
The relation of $1/T_1T\propto N(E_{\rm F})^2$ enables us to deduce the density of states at the Fermi surface $N(E_{\rm F})$. 
As shown in Fig. \ref{fig:T1}(b), it was found that (111)-thick film with higher effective carrier density has larger $N(E_{\rm F})$ than (100)-thin film with lower carrier density does, suggesting  that the enhancement of $T_{\rm c}$ in (111)-thick film primarily relates to the increase of $N(E_{\rm F})$ that is induced by the effective carrier doping by the substitution at B(1) site. 
Note that their $T_{\rm c}$s were not precisely predicted only by increase of $N(E_{\rm F})$ in McMillan's formula\cite{McMillan} when assuming the same Debye frequency and electron-phonon interaction, suggesting that it is necessary to consider the doping dependence of these parameters to describe its $T_c$.

\begin{figure}[htbp]
\begin{center}
\includegraphics[width=0.85\linewidth]{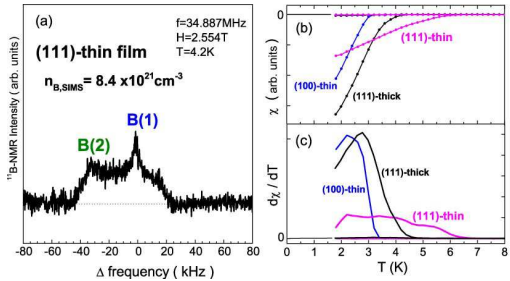}
\end{center}
\caption{(color online) (a) NMR spectra, (b) susceptibility and (c) its $T$-differential for (111)-thin film, together with the others. The diamagnetism in (111)-thin film gradually occurs over the broad $T$-range between 2 K and 7 K. }
\label{fig:thinfilm}
\end{figure}

Finally we comment on the result in the (111)-thin film ($n_{\rm B,SIMS}=8.4\times10^{21}$cm$^{-3}$) with highest-$T_{\rm c}$  of 7 K in this work. 
Although $T_{\rm c}$ is two times higher than that for (100)-thin film in spite of very similar $n_{\rm B,SIMS}$, the apparent differences has not been observed in their spectral shape, as shown in Fig. \ref{fig:thinfilm}(a). 
It must be noted that the (111)-thin films tend to exhibit broader SC transition than the (100) films does(see Refs.\cite{Takano,Bustarret,Umezawa}). 
Indeed, the diamagnetic response for this (111)-thin film appears over the broad $T$-range between 2 K and 7 K as indicated in Figs. \ref{fig:thinfilm}(b) and \ref{fig:thinfilm}(c). 
Recently a possible distribution of $T_{\rm c}$ in the (111)-thin films has been thoroughly investigated by grinding the (111)-thin film from surface and/or bottom, and as a result, it is revealed that $T_{\rm c}$  becomes higher close to the surface by means of transport measurement\cite{Takenouchi}.
It suggests the spatially distribution of $T_{\rm c}$ for the (111)-thin films over a range in $T_{\rm c}$=2-7 K.
The intensity of NMR spectra reflects the spatially-averaged B(1) and B(2) densities for all domains with different $T_{\rm c}$'s. 
In this situation, we tentatively supposed $T_{\rm c}\sim$4 K as a spatially-averaged value, determined by a center of the broad peak in $d\chi/dT$ {\it vs} $T$ curve in Fig. \ref{fig:thinfilm}(c). 
Using the $n_{\rm B(1)}\sim 1\times10^{21}$cm$^{-3}$ from the spectrum analysis, it is reasonably plotted on the same curve in Fig. \ref{fig:Phase}, suggesting the spatially-averaged $T_{\rm c}$ in the (111)-thin film seems to be related to $n_{\rm B(1)}$ as well.
In addition, $1/T_1T$ in (111)-thin film is also similar to that in (100)-thin film, as shown in Fig. \ref{fig:T1}, suggesting that both $N(E_{\rm F})$ and $n_{\rm B(1)}$ are also similar with each other. 
%**************  Fig.6 *****************************************************
%****************************************************************************

However, there still remains a question why thin films includes more B(2) than thick film does. 
With respect to B+H complex, there are possible H positions such as bond-center, back-bond positions of boron-carbon, $C_{2v}$ symmetry site\cite{Yoshida,Oguchi}, B+H$_{\rm n}$ complexes\cite{Goss} as suggested by the band calculations.
Recently Oguchi calculated the EFG for these states and obtained a similar value of $\nu_Q$ to our experiment for the case of the bond-center position of B-C\cite{Oguchi}.
Either the relaxation process of the lattice constant in the thick films or a non-equilibrium process of MP-CVD method may control the formation of B+H complex. 
The annealing has been tried to expel the hydrogen forming the B+H complex, but it has not been succeeded yet.

%* summary **

In conclusion, the present $^{11}$B-NMR study on the B-doped diamond revealed that borons are doped into the substitutional B(1) site as single occupation and as lower symmetric B(2) sites forming boron+hydrogen (B+H) complex. This result has enabled us to extract the scaling behavior between $T_{\rm c}$  and the B(1) concentration $n_{\rm B(1)}$. 
A clear evidence is presented that the charge carriers introduced by B(1) substitution are responsible for the superconductivity, whereas the charge neutral B(2) sites substituted as B+H complex does not offer the carriers effectively. 
A route to make $T_{\rm c}$  higher in the boron doped diamond is to dissociate the hydrogen at B(2) site by annealing, and to find how it can be possible to make the boron concentration at the substitutional B(1) site increase, keeping the diamond structure. 
This may also help developing diamond-based devices that make use of the unique properties of diamond. 

%* acknowledgement **

The authors would like to thank H. Katayama-Yoshida, T. Yokoya for their valuable discussions and comments.  This work was supported by Grant-in-Aid for Creative Scientific Research (15GS0213) from the Ministry of Education, Culture, Sports, Science and Technology (MEXT) and the 21st Century COE Program (G18) by Japan Society of the Promotion of Science (JSPS). 

%::::::::::::::::bibliography::::::::::::::::::::::::::::::::::::::::::::::::
 %:::::::::::::::::::::::::::::::::::::::::::::::::::::::

\begin{references}
\bibitem{parameters} The used parameters are $K\sim-10$ppm, $\nu_Q(1)\sim0$ for B(1) site and $\nu_Q(2)\sim1.6$MHz for B(2) site with some broadening in frequency.
\bibitem{broadpeak} The B occupation at the interstitial site is excluded  because one B atom in this site kills three holes in the hole-doped diamond. It is also expected to be unstable by the band calculation\cite{Yoshida}.
\end{references}
\end{document}